\documentclass[prl,letterpaper,twocolumn,showpacs,superscriptaddress,amsmath,amsfonts,amssymb,preprintnumbers]{revtex4}
\pdfoutput=1
\usepackage[T1]{fontenc}\usepackage[latin1]{inputenc}
\usepackage{dcolumn,graphicx,color,hvfloat,booktabs,microtype,afterpage} 
\usepackage[charter]{mathdesign}
\renewcommand{\figurename}{Fig.}
\renewcommand{\tablename}{Table}
\makeatletter\renewcommand{\fnum@figure}[1]{\figurename~\thefigure~(color online).}\makeatother
\makeatletter\renewcommand{\fnum@table}[1]{\tablename~\thetable.}\makeatother
\definecolor{Red}{rgb}{0.8,0,0.0}

\usepackage[colorlinks,plainpages=false,linkcolor=black,urlcolor=blue,citecolor=black,pdfpagemode=UseNone,pdfstartview=FitBH]{hyperref}

\begin{document} \pagestyle{plain}

\title{Electronic structure and nesting-driven enhancement of the RKKY interaction\\at the magnetic ordering propagation
vector in Gd$_2$PdSi$_3$ and Tb$_2$PdSi$_3$}

\author{D.\,S.\,Inosov}
\affiliation{Leibniz Institut für Festkörper- und Werkstoffforschung Dresden, D-01171 Dresden, Germany.}
\affiliation{Max-Planck-Institute for Solid State Research, Heisenbergstraße~1, D-70569 Stuttgart, Germany.}
\author{D.\,V.~Evtushinsky}\author{A.\,Koitzsch}\author{V.\,B.\,Zabolotnyy}\author{S.\,V.~Borisenko}
\affiliation{Leibniz Institut für Festkörper- und Werkstoffforschung Dresden, D-01171 Dresden, Germany.}
\author{A.\,A.\,Kordyuk}
\affiliation{Leibniz Institut für Festkörper- und Werkstoffforschung Dresden, D-01171 Dresden, Germany.}
\affiliation{Institute of Metal Physics of National Academy of Sciences of Ukraine, 03142 Kyiv, Ukraine.}
\author{M.\,Frontzek}\author{M.\,Loewenhaupt}
\affiliation{Institut für Festkörperphysik, Technische Universität Dresden, D-01062 Dresden, Germany.}
\author{W.~Löser}\author{I.\,Mazilu}\author{H.\,Bitterlich}\author{G.\,Behr}
\affiliation{Leibniz Institut für Festkörper- und Werkstoffforschung Dresden, D-01171 Dresden, Germany.}
\author{J.-U.\,Hoffmann}
\affiliation{Helmholtz-Zentrum Berlin für Materialien und Energie GmbH, D-14109 Berlin, Germany.}
\author{R.\,Follath}
\affiliation{BESSY GmbH, Albert-Einstein-Strasse 15, 12489 Berlin, Germany.}
\author{B.\,Büchner}
\affiliation{Leibniz Institut für Festkörper- und Werkstoffforschung Dresden, D-01171 Dresden, Germany.}

\keywords{intermetallic compounds, heavy fermions, Fermi surface, electronic structure, photoemission spectra}

\pacs{71.20.Lp 71.27.+a 79.60.-i 71.18.+y 74.25.Jb}


\begin{abstract}

\noindent We present first-time measurements of the Fermi surface and low-energy electronic structure of intermetallic
compounds Gd$_2$PdSi$_3$ and Tb$_2$PdSi$_3$ by means of angle-resolved photoelectron spectroscopy (ARPES). Both
materials possess a flower-like Fermi surface consisting of an electron barrel at the $\Gamma$ point surrounded by
spindle-shaped electron pockets originating from the same band. The band bottom of both features lies at 0.5\,eV below
the Fermi level. From the experimentally measured band structure, we estimate the momentum-dependent RKKY coupling
strength and demonstrate that it is peaked at the $\frac{1}{\text{\raisebox{1pt}{2}}}\Gamma$K wave vector. Comparison
with neutron diffraction data from the same crystals shows perfect agreement of this vector with the propagation vector
of the low-temperature in-plane magnetic order, thereby demonstrating the decisive role of the Fermi surface geometry in
explaining the complex magnetically ordered ground state of ternary rare earth silicides.

\end{abstract}

\maketitle

\noindent Ternary rare earth silicides with hexagonal crystal structure of the form R$_2$PdSi$_3$, where R is a rare
earth atom, are known to exhibit complex magnetic behavior \cite{KotsanidisYakinthos90, MallikSampathkumaran98,
MallikSampathkumaran98ssc, SahaSugawara99, MallikSampathkumaran99, SzytulaHofmann99, FrontzekKreyssig07, SahaSugawara00,
MajumdarSampathkumaran00, SampathkumaranBitterlich02, LiNimori03, PauloseSampathkumaran03, FrontzekKreyssig06,
CollectionMagnetic} due to a delicate competition of the Ruderman-Kittel-Kasuya-Yosida (RKKY) interaction
\cite{CollectionRKKY} and the Kondo effect, which are comparable in magnitude \cite{MallikSampathkumaran98,
SahaSugawara00}. Such interplay determines many unusual magnetic, thermal, and transport properties, which stimulate
unceasing interest to these materials during the last two decades: large negative magnetoresistance
\cite{SahaSugawara99}, quasi-low-dimensional magnetism and spin-glass-like behavior \cite{PauloseSampathkumaran03,
LiNimori03}, highly anisotropic ac susceptibility \cite{PauloseSampathkumaran03}, magnetocaloric effect
\cite{MajumdarSampathkumaran00}, thermoelectric power \cite{SahaSugawara00}, and Hall coefficient \cite{SahaSugawara00}.

Most of the R$_2$PdSi$_3$ compounds order magnetically at low temperatures, somewhat below the Kondo minimum in the
resistivity \cite{MallikSampathkumaran98, MallikSampathkumaran98ssc, SahaSugawara99, SzytulaHofmann99,
SampathkumaranBitterlich02}. The exact type of such ordering strongly depends on the material and can be rather
complicated \cite{SzytulaHofmann99, FrontzekKreyssig07}. The corresponding Neel temperature $T_\text{N}$ reaches maximum
for the Gd ($T_\text{N}$=21.0\,K \cite{KotsanidisYakinthos90}) and Tb ($T_\text{N}$=23.6\,K \cite{FrontzekKreyssig06})
compounds, which we have chosen as the subject of the present study. The RKKY exchange interaction that essentially
determines their magnetic properties is mediated by the conductance electrons, and therefore any reasonable description
of the corresponding physics is impossible without the knowledge of the Fermi surface and the low-energy electronic
structure. According to our recent study \cite{EvtushinskyKordyuk08}, the sign reversal of the Hall effect observed at
low temperatures in R$_2$PdSi$_3$ \cite{MallikSampathkumaran99} might be an indication of the opening of the pseudogap
at some portions of the Fermi surface, as was also suggested earlier by resistivity measurements
\cite{MallikSampathkumaran98}, again emphasizing the importance of studying the electronic structure of these materials.
Nevertheless, though the single crystals of Gd$_2$PdSi$_3$ are available since nearly a decade \cite{SahaSugawara99},
the fermiology and the underlying band structure are still not known for any of the R$_2$PdSi$_3$ compounds neither from
momentum-resolved measurements nor from band structure calculations. Earlier photoemission experiments
\cite{ChaikaIonov01, SzytulaJezierski03} were performed only on polycrystalline samples and could not therefore shed
light on the dispersion of conductance electrons and the Fermi surface geometry, but have revealed that the density of
states at the Fermi level is likely to be dominated by the 5d states of the rare earth atoms.

\begin{figure}[b]\vspace{-0.8em}
        \includegraphics[width=0.60\columnwidth]{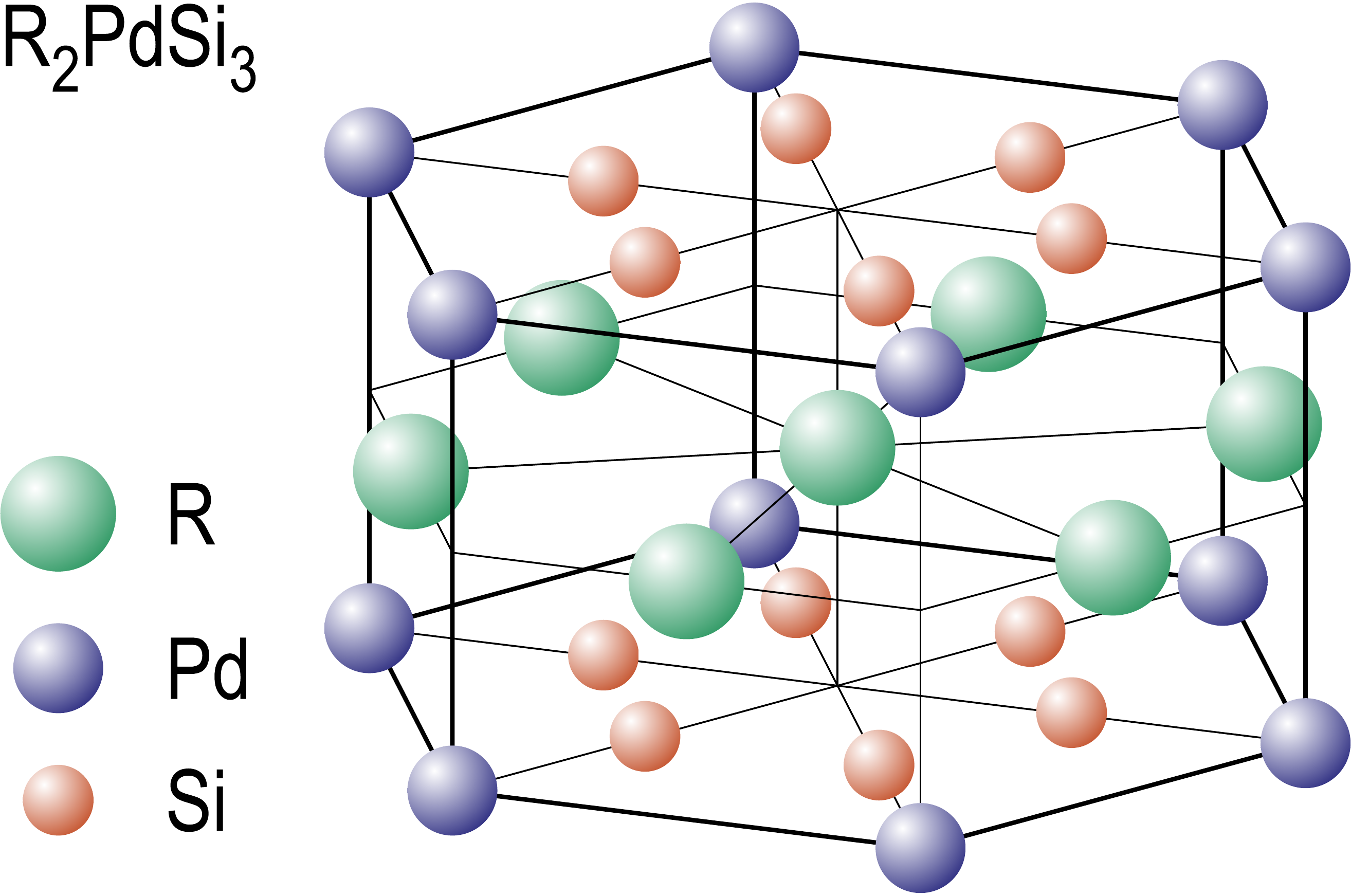}\vspace{-0.3em}
        \caption{Crystal structure of R$_2$PdSi$_3$ after Ref.\,\onlinecite{PauloseSampathkumaran03}\,\&\,\onlinecite{ChaikaIonov01}.\vspace{-1.5em}}\label{Fig:CrystalStructure}
\end{figure}

Here we report an angle-resolved photoelectron spectroscopy (ARPES) investigation of the low-energy electronic structure
performed on the single crystals of Gd$_2$PdSi$_3$ and Tb$_2$PdSi$_3$, whose crystal structure is illustrated by Fig.\,\ref{Fig:CrystalStructure}. We show that the Fermi surface in both compounds
has a flower-like shape with the dominant nesting vector that coincides with the propagation vector of the
low-temperature magnetic ordering, as we will see from the comparison of our ARPES data and neutron diffraction patterns
measured from the same single crystals. This observation offers a simple explanation for the complex magnetic ordering,
namely that it is determined by the enhanced RKKY coupling at the nesting wave vector of the normal-state Fermi surface.

The single crystals for the present study were grown by the floating-zone method from stoichiometric polycrystalline
feed rods \cite{CrystalGrowth} and cleaved \textit{in situ} perpendicular to the (001) direction immediately before
measurement in ultra-high vacuum of 10$^{-10}$\,mbar. For a detailed description of our experimental geometry see
Ref.\,\onlinecite{InosovSchuster08}. The measurements were performed with 150\,eV photons\,---\,the excitation energy
that was found to yield maximal photocurrent.

\begin{figure}[t]\vspace{-0.3em}
        \includegraphics[width=\columnwidth]{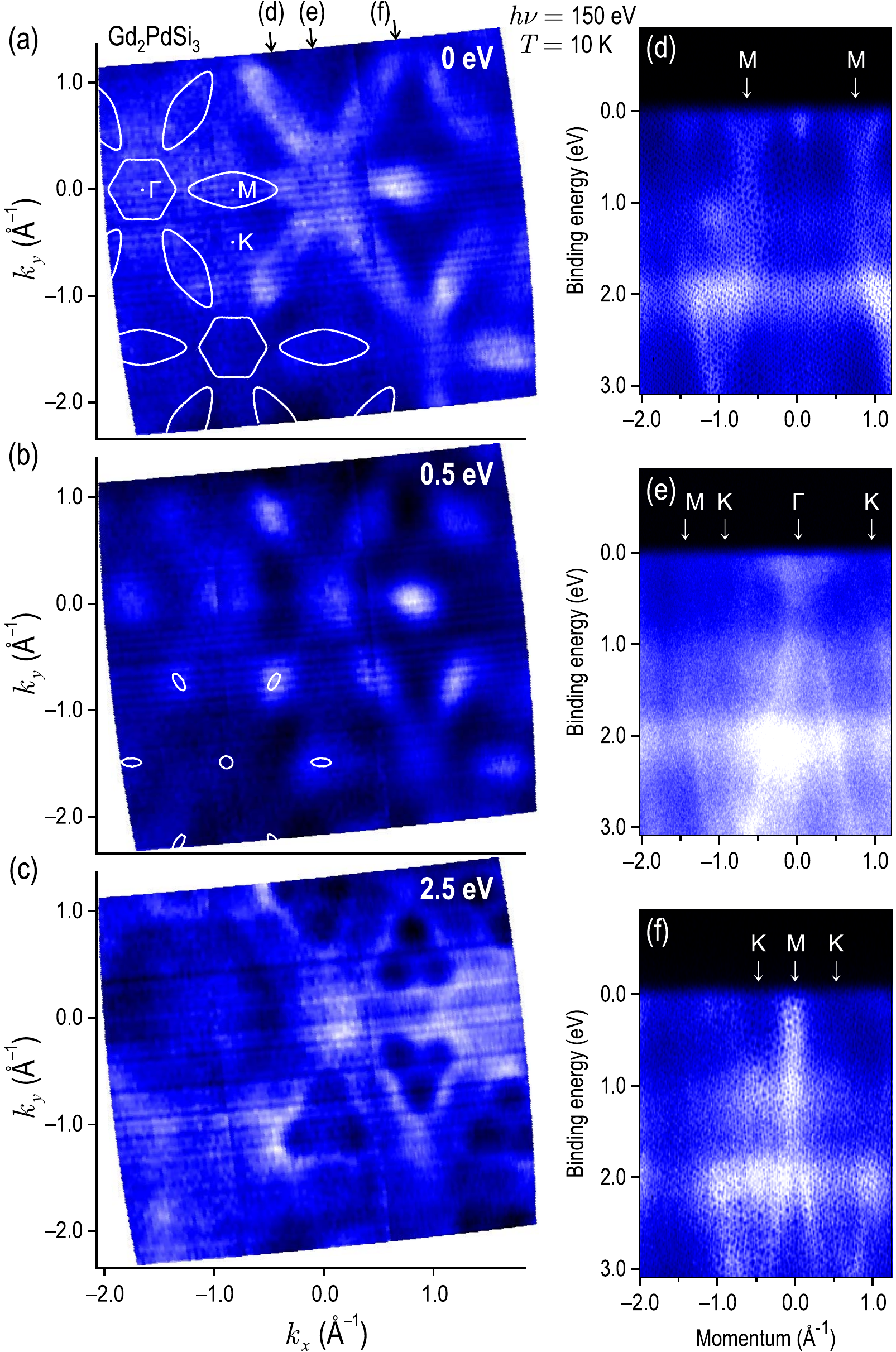}\vspace{-0.3em}
        \caption{Fermi surface and the underlying electronic structure of Gd$_2$PdSi$_3$. Lighter colors represent
        higher photoemission intensity. Panels (a)\,--\,(c) show constant energy cuts taken at the Fermi level, at 0.5
        and 2.5\,eV binding energy respectively. Panels (d)\,--\,(f) show energy-momentum cuts as indicated by arrows in
        panel (a). The solid white contours in panels (a) and (b) represent the tight-binding fit to the data.\vspace{-1.5em}}\label{Fig:Gd2PdSi3}
\end{figure}

\begin{figure}[b]\vspace{-0.8em}
        \includegraphics[width=\columnwidth]{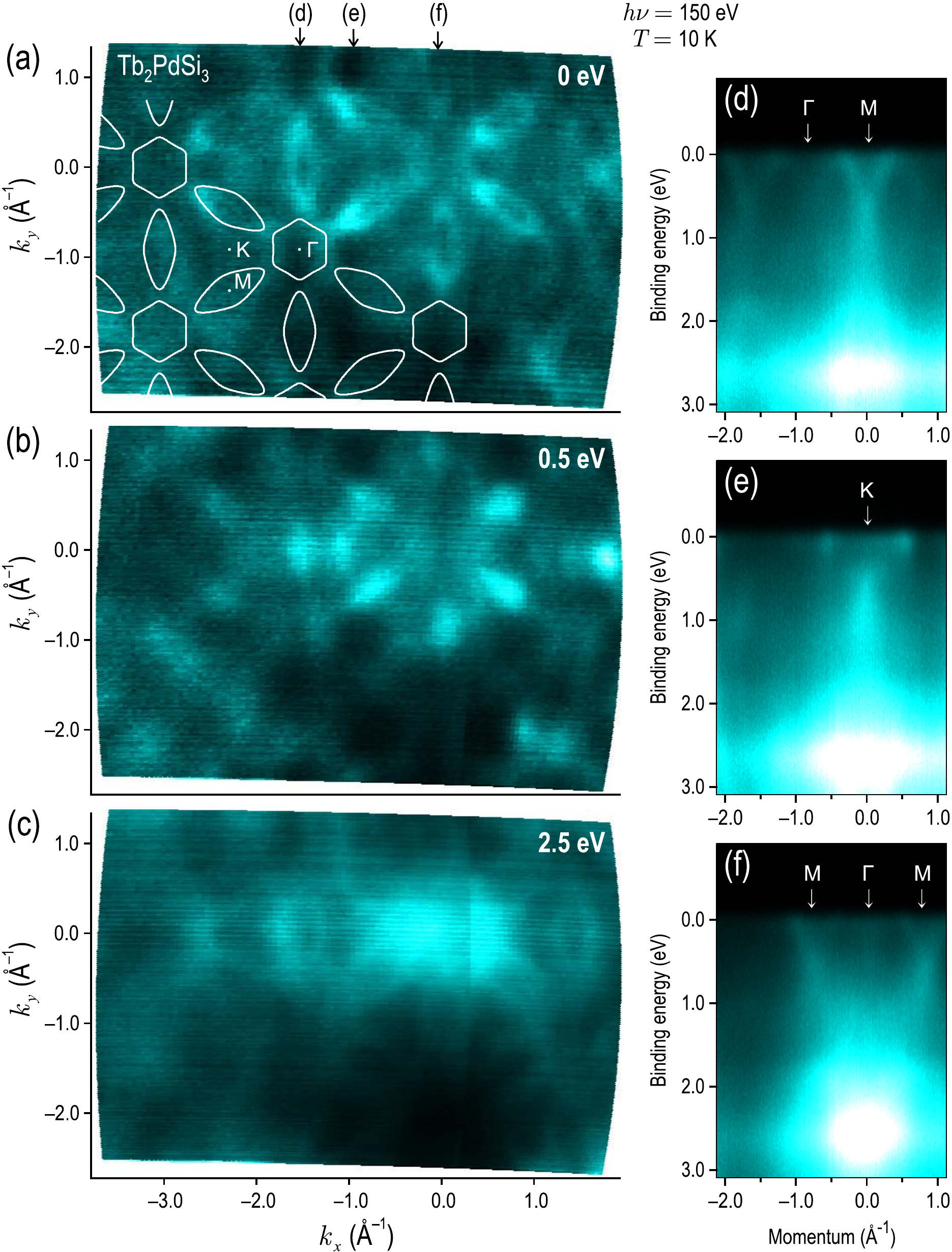}\vspace{-0.3em}
        \caption{Same as Fig.\,\ref{Fig:Gd2PdSi3}, but for Tb$_2$PdSi$_3$. Note the similar electronic structure of Gd
        and Tb compounds near the Fermi level and the differences at higher binding energies, where Tb\,4f states start
        to contribute.\vspace{-1.2em}}\label{Fig:Tb2PdSi3}
\end{figure}

Fig.\,\ref{Fig:Gd2PdSi3} shows several constant-energy cuts and energy-momentum cuts representing
electron dispersions in Gd$_2$PdSi$_3$, measured at low temperature ($T$\,=\,10\,K). Panel (a) shows the Fermi
surface, which consists of an electron barrel at the $\Gamma$ point surrounded by spindle-shaped electron pockets
originating from the same band, as sketched by the white lines. Though the $\Gamma$-barrel is not well distinguishable
in panel (a) due to its low intensity, it can be clearly seen in panel (e), which represents a K--$\Gamma$--K
energy-momentum cut. The band bottoms of both $\Gamma$- and M-centered barrels lie at 0.5\,eV below the Fermi level, as
follows from panels (d) and (e), where similar point-like intensity blobs can be seen at both $\Gamma$ and M points (b).
At yet higher binding energies, the $\Gamma$- and M-centered features increase again in size, resulting in a fancy
trefoil-like structure at 2.5\,eV (c).

The Fermi surface of Tb$_2$PdSi$_3$, shown in Fig.\,\ref{Fig:Tb2PdSi3}\,(a), is very similar, though the
$\Gamma$-centered barrel is much less intense in the photoemission spectra (b). Nevertheless, it can still be recognized
in panel (f) of the same figure, which also shows that both electron-like barrels extend down to 0.5\,eV binding energy
exactly as in the Gd compound. At higher binding energies, the electronic structures of the two compounds start to
differ (c). Instead of the trefoil-like structure, an intense feature at the $\Gamma$ point is observed in
Tb$_2$PdSi$_3$ near 2.5\,eV, which can be assigned to the lowest line of the
\hbox{Tb\,4f$\kern1pt^8$\hspace{0.5pt}$\rightarrow$\,4f$\kern.7pt^7$} final state multiplet \cite{Gerken82,
ChaikaIonov01, SzytulaJezierski03}. The corresponding \hbox{4f$\kern1pt^7$\hspace{0.5pt}$\rightarrow$\,4f$\kern.7pt^6$}
multiplet in Gd is located at higher binding energies and is therefore not observed in our spectra.

\begin{figure}[t]
        \includegraphics[width=\columnwidth]{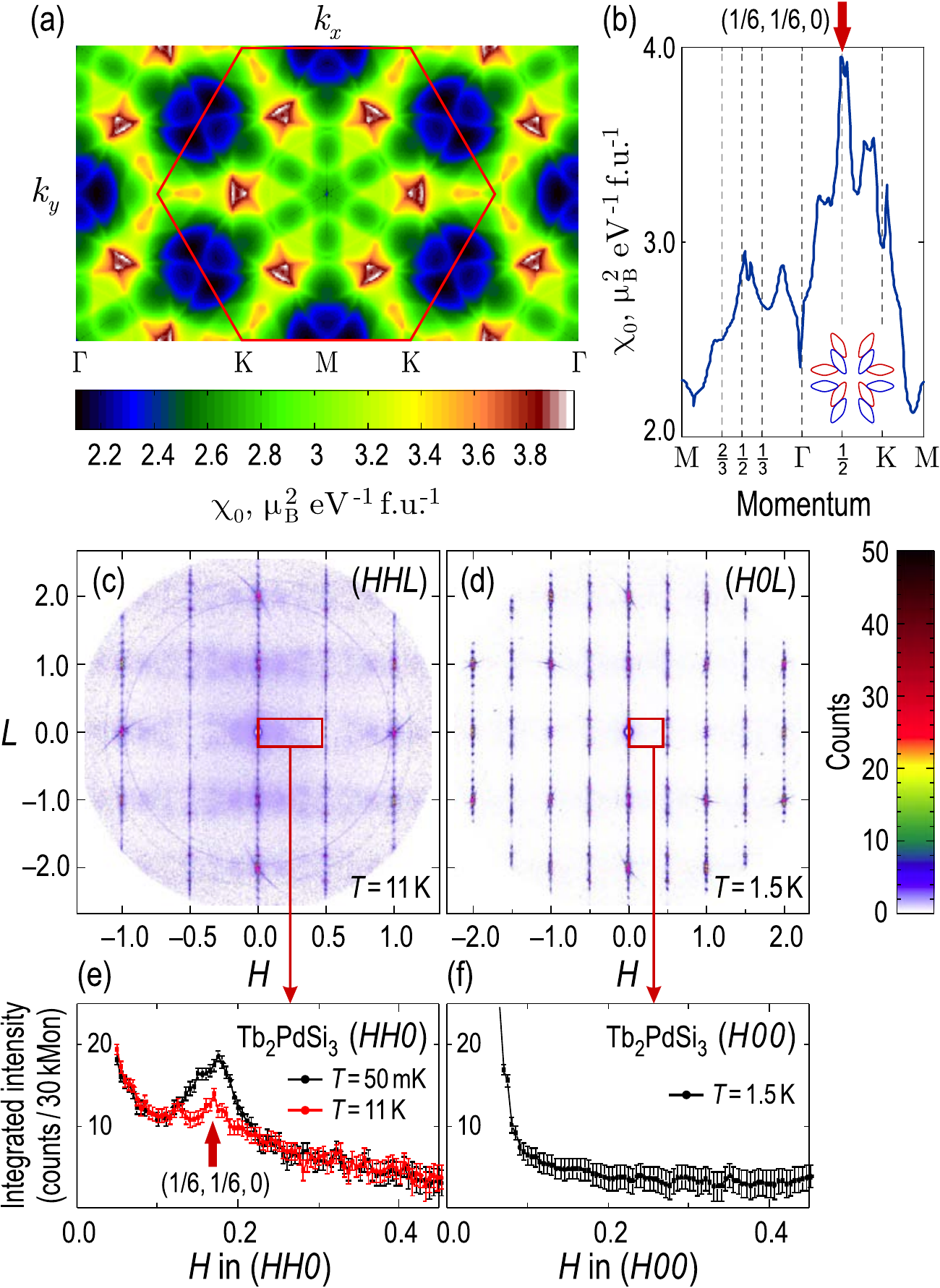}
        \caption{Nesting properties of the Tb(Gd)$_2$PdSi$_3$ Fermi surface. (a)~Real part of the Lindhard function at
        $\omega\rightarrow0$ as a function of momentum. The hexagon marks the Brillouin zone boundary.
        (b)~Corresponding profile along high-symmetry directions, with the dominant nesting vector marked by the arrow.
        The same vectors can be seen in panel (a) as white spots. The inset shows the external tangency of the
        spindle-shaped pockets responsible for the nesting peak at the $\frac{1}{2}\Gamma$K wave vector.
        (c)~and~(d)~Neutron diffraction patterns measured from Tb$_2$PdSi$_3$ single crystals in the $(HHL)$ and $(H0L)$
        planes at 11\,K and 1.5\,K respectively. (e)~and~(f)~Corresponding intensity profiles along $H$, integrated
        within the rectangles shown in panels (c) and (d). The arrow in panel (e) marks that position of the diffraction
        peak that coincides with the nesting peak in panel (b).\vspace{-1.5em}}\label{Fig:NestingR2PdSi3}
\end{figure}

Now let us look in more detail at the Fermi surface geometry, as it is shown in
Fig.\,\ref{Fig:Gd2PdSi3}\,and\,\ref{Fig:Tb2PdSi3}\,(a), in order to investigate how it affects the momentum-dependent
strength of the RKKY coupling and thereby influences the complex magnetic ordering structure that sets in at low
temperatures. It is known that in the linear response assumption the coupling constant of the RKKY interaction is
determined by the itinerant spin susceptibility of the material (Lindhard function) \cite{CollectionRKKY}. From a
tight-binding fit to the experimentally measured band structure of R$_2$PdSi$_3$, following a procedure similar to that
used in Ref.\,\onlinecite{InosovSuscept}, we have calculated the Lindhard function in the static limit
($\omega\rightarrow0$), which is shown in Fig.\,\ref{Fig:NestingR2PdSi3}~(a)~and~(b). One sees a strong peak at the
$\frac{1}{\text{\raisebox{1pt}{2}}}\Gamma$K wave vector that originates from the perfect nesting of spindle-shaped
pockets as sketched in the inset. This peak indicates that the indirect RKKY exchange interaction should be strongly
enhanced at the $\bigl(\frac{1}{\text{\raisebox{0.5pt}{6}}}\,\frac{1}{\text{\raisebox{0.5pt}{6}}}\,0\bigr)$ wave vector
(i.e. half way from the $\Gamma$ point towards the corner of the hexagonal Brillouin zone). If one compares this result
with the low-temperature neutron diffraction data \cite[Fig.\,2\,and\,3]{FrontzekKreyssig07}, one sees that the magnetic
structure in Tb$_2$PdSi$_3$ indeed sets in with the
$\bigl(\frac{1}{\text{\raisebox{0.5pt}{6}}}\,\frac{1}{\text{\raisebox{0.5pt}{6}}}\,0\bigr)$ in-plane component of the
propagation vector.

In Fig.\,\ref{Fig:NestingR2PdSi3}\,(c)\,--\,(f) we show neutron diffraction data obtained from the same single crystals
of Tb$_2$PdSi$_3$ \cite{NoteGd} as our photoemission measurements. Panel (c) shows data in the reciprocal $(HHL)$ plane
measured at 11\,K, whereas panel (d) shows data in the $(H0L)$ plane. Again, we see a well-defined peak at
$H=\frac{1}{\text{\raisebox{0.5pt}{6}}}$ along the (110) direction (Brillouin zone diagonal) already at 11\,K, which
gets even stronger upon cooling (e), whereas no peaks along (100) direction are observed down to 1.5\,K (f). This
observation confirms that the low-temperature short-range magnetic ordering of the localized Tb\,4f electrons is
characterized by the $\bigl(\frac{1}{\text{\raisebox{0.5pt}{6}}}\,\frac{1}{\text{\raisebox{0.5pt}{6}}}\,0\bigr)$
propagation vector, as dictated by the nesting-driven enhancement of the RKKY interaction at this wave vector. The
intense peaks observed at $H=\pm0.5$ and $H=0$ in both diffraction patterns originate from the long-range magnetic order
predetermined by the crystallographic superstructure, which is driven mainly by the antiferromagnetic exchange
interaction along the $L$-direction. The corresponding diffraction reflections are therefore irrelevant to our
discussion, since the out-of-plane component is inaccessible in our present ARPES measurements. Our recent neutron
diffraction studies (not shown) have revealed magnetic order with the same in-plane propagation vector of
$\bigl(\frac{1}{\text{\raisebox{0.5pt}{6}}}\,\frac{1}{\text{\raisebox{0.5pt}{6}}}\,0\bigr)$ also in Dy$_2$PdSi$_3$
($T_\text{N}=8.2$\,K \cite{FrontzekKreyssig06}).

It is appropriate to mention here that magnetic structures with the
$\bigl(\frac{1}{\text{\raisebox{0.5pt}{6}}}\,\frac{1}{\text{\raisebox{0.5pt}{6}}}\,0\bigr)$ propagation vector were
proposed in other rare earth intermetallics with magnetically frustrated triangular lattices, such as ZrNiAl-type
systems, where indirect RKKY exchange also plays a role \cite{GondekSzytula07}. This might motivate future studies of
the electronic structure in intermetallic compounds. Ref.\,\onlinecite{GondekSzytula07} also proposes an explicit
arrangement of magnetic moments on a triangular lattice that corresponds to the
$\bigl(\frac{1}{\text{\raisebox{0.5pt}{6}}}\,\frac{1}{\text{\raisebox{0.5pt}{6}}}\,0\bigr)$ kind of ordering, which
might turn out to be relevant for the R$_2$PdSi$_3$ systems as well.

To conclude, we have investigated the low-energy electronic structure and Fermi surface geometry of two ternary rare
earth silicides that show highest Neel temperatures among the R$_2$PdSi$_3$ series of compounds. We have shown that both
compounds possess similar Fermi surfaces with a strong nesting vector that enhances indirect RKKY exchange interaction
at the $\bigl(\frac{1}{\text{\raisebox{0.5pt}{6}}}\,\frac{1}{\text{\raisebox{0.5pt}{6}}}\,0\bigr)$ wave vector, in
perfect agreement with the in-plane propagation vector of the low-temperature magnetic ordering observed in our neutron
diffraction measurements on the same single crystals. Therefore we conclude that the low-temperature magnetic ordering
originates from the RKKY interaction, mediated by itinerant Tb\,5d electrons at the Fermi level, and is driven by the
enhanced itinerant spin susceptibility at the magnetic propagation vector, which can be understood within a simple Fermi
surface nesting picture.

This~project~is~part~of~the~Forschergruppe~FOR538 and is partially supported by the DFG under Grants No.~KN393/4 and
BO1912/2-1, as well as DFG research project SFB\,463. ARPES experiments were performed using the $1^3$-ARPES end station
at the UE112-lowE PGMa beamline of the Berliner Elektronenspeicherring-Gesellschaft für Synchrotron Strahlung m.b.H.
(BESSY). Authors gratefully acknowledge the scientific and technical support from the team of the E2 diffractometer at
Helmholtz-Zentrum Berlin für Materialien und Energie GmbH, where neutron scattering experiments were performed. We also
thank R.\,Hübel, S.\,Leger, and R.\,Schönfelder for technical support and N.\,Wizent for assistance with sample
preparation.


\begin{thebibliography}{10}

\bibitem{SahaSugawara00} S.\,R.\,Saha, H.\,Sugawara, T.~D.\,Matsuda, Y.~Aoki, H.\,Sato, and E.\,V.~Sampathkumaran, \prb \textbf{62}, 425 (2000).
\bibitem{MallikSampathkumaran98} R.\,Mallik, E.\,V.~Sampathkumaran, M.\,Strecker, and G.\,Wort\-mann, Europhys. Lett. \textbf{41}, 315 (1998).
\bibitem{SahaSugawara99} S.\,R.\,Saha, H.\,Sugawara, T.~D.\,Matsuda, H.\,Sato, R.\,Mallik, and E.\,V.~Sampathkumaran, \prb \textbf{60}, 12162 (1999).
\bibitem{SampathkumaranBitterlich02} E.\,V.~Sampathkumaran, H.\,Bitterlich, K.\,K.\,Iyer, W.\,Löser, and G.\,Behr, \prb \textbf{66}, 052409 (2002).
\bibitem{MallikSampathkumaran98ssc} R.\,Mallik, E.\,V.~Sampathkumaran, and P.~L.\,Paulose, Sol. State Comm. \textbf{106}, 169 (1998).
\bibitem{SzytulaHofmann99} A.\,Szytula, M.\,Hofmann, B.\,Penc, M.\,Slaski, S.\,Majumdar, E.\,V.~Sampathkumaran, and A.\,Zygmunt, J. Magn. Magn. Mater. \textbf{202}, 365 (1999).
\bibitem{PauloseSampathkumaran03} P.~L.\,Paulose, E.\,V.~Sampathkumaran, H.\,Bitterlich, G.\,Behr, and W. Löser, \prb \textbf{67}, 212401 (2003).
\bibitem{LiNimori03} D.\,X.\,Li, S.\,Nimori, Y.~Shiokawa, Y.~Haga, E.\,Yamamoto, and Y.~Onuki, \prb \textbf{68}, 012413 (2003).
\bibitem{MajumdarSampathkumaran00} S.\,Majumdar, E.\,V.~Sampathkumaran, P.~L.\,Paulose, H.\,Bitter\-lich, W.\,Löser, and G.\,Behr, \prb \textbf{62}, 14207 (2000).
\bibitem{FrontzekKreyssig07} M.\,Frontzek, A.\,Kreyssig, M.\,Doerr, A.\,Schneidewind, J.-U.\,Hoffmann, and M. Loewenhaupt, J.\,Phys.: Cond.\,Matt. \textbf{19}, 145276 (2007).
\bibitem{KotsanidisYakinthos90} P.~A.\,Kotsanidis, J.\,K.\,Yakinthos, and E.\,Gamari-Seale, J.~Magn. Magn. Mater. \textbf{87}, 199 (1990).
\bibitem{FrontzekKreyssig06} M.\,Frontzek, A.\,Kreyssig, M.\,Doerr, M.\,Rotter, G.\,Behr, W.\,Löser, I.\,Mazilu, and M. Loewenhaupt, J.~Magn. Magn. Mater. \textbf{301}, 398 (2006).
\bibitem{MallikSampathkumaran99} R.\,Mallik, E.\,V.~Sampathkumaran, P.~L.\,Paulose, H.\,Sugawara, and H.\,Sato, Physica\,B \textbf{259--261}, 892 (1999).
\bibitem{CollectionMagnetic} R.\,Mallik and E.\,V.~Sampathkumaran, J.~Magn. Magn. Mater. \textbf{164}, L13 (1996);
   E.\,V.~Sampathkumaran, I.\,Das, R.\,Rawat, and S.\,Majumdar, Appl. Phys. Lett. \textbf{77}, 418 (2000);
   S.\,Majumdar, H.\,Bitterlich, G.\,Behr, W.\,Löser, P.~L.\,Paulose, and E.\,V.~Sampathkumaran, \prb \textbf{64}, 012418 (2001);
   M.\,Frontzek, A.\,Kreyssig, M.\,Doerr, J.-U.\,Hoffman, D.\,Hohlwein, H.\,Bitterlich, G.\,Behr, and M. Loewenhaupt, Physica\,B \textbf{350}, e187 (2004);
   K.\,K.\,Iyer, P.~L.\,Paulose, E.\,V.~Sampathkumaran, M.\,Frontzek, A.\,Kreyssig, M.\,Doerr, M.\,Loewenhaupt, I.\,Mazilu, G.\,Behr, W.\,Löser, Physica\,B \textbf{255}, 158 (2005).
\bibitem{CollectionRKKY} M.\,A.\,Ruderman and C.\,Kittel, Phys.~Rev.~\textbf{96}, 99 (1954);
   T.~Kasuya, Prog.~Theor.~Phys. (Kyoto) \textbf{16}, 45 (1956);
   K.\,Yosida, Phys.~Rev.~\textbf{106}, 893 (1957);
   C.\,Kittel, in \textit{Solid State Physics}, edited by F.~Zeitz, D.\,Turnbull, and H.\,Ehreinreich (Academic, New York, 1968), vol. 22, p.\,1;
   Y.~Yafet, \prb \textbf{36}, 3948 (1987);
   J.\,G.\,Kim, E.\,K.\,Lee, and S.\,Lee, \prb \textbf{54} (1996);
   D.\,N.\,Aristov, \prb \textbf{55}, 8064 (1997);
   V.~I.\,Litvinov and V.~K.\,Dugaev, \prb \textbf{58}, 3584 (1998).
\bibitem{EvtushinskyKordyuk08} D.\,V.~Evtushinsky, S.\,V.~Borisenko, A.\,A.\,Kordyuk, V.~B.\,Zabo\-lotnyy, D.\,S.\,Inosov, B.\,Büchner, H.\,Berger, L.\,Patthey, and R.\,Follath, \prl \textbf{100}, 236402 (2008).
\bibitem{ChaikaIonov01} A.\,N.\,Chaika, A.\,M.\,Ionov, M.\,Busse, S.\,L.\,Molodtsov, S.\,Maj\-um\-dar, G.\,Behr, E.\,V.~Sampathkumaran, W.\,Schneider, and C.\,Laubschat, \prb \textbf{64}, 125121 (2001);
\bibitem{SzytulaJezierski03} A.\,Szytu{\l}a, A.\,Jezierski, and B.\,Penc, Physica\,B \textbf{327}, 171 (2003).
\bibitem{CrystalGrowth} G.\,Graw, H.\,Bitterlich, W.\,Löser, G.\,Behr, J.\,Fink, and L.\,Schultz, J.~Alloys~Compd. \textbf{308}, 193 (2000);
   G.\,Behr, W.~Löser, D.\,Souptel, G.\,Fuchs, I.\,Mazilu, C.\,Cao, A.\,Köhler, L.\,Schultz, and B.\,Büchner, J.~Cryst. Growth \textbf{310}, 2268 (2008).
\bibitem{InosovSchuster08} D.\,S.\,Inosov, R.\,Schuster, A.\,A.\,Kordyuk, J.\,Fink, S.\,V.~Bori\-sen\-ko, V.~B.\,Zabolotnyy, D.\,V.~Evtushinsky, M.\,Knupfer, B.\,Büchner, R.\,Follath, and H.\,Berger, \prb \textbf{77}, 212504 (2008).
\bibitem{Gerken82} F.~Gerken, J.\,Phys.\,F: Met.\,Phys. \textbf{13}, 703 (1982).
\bibitem{InosovSuscept} D.~S.~Inosov, S.~V.~Borisenko, I.~Eremin, A.~A.~Kordyuk, V.~B.~Zabolotnyy, J.\,Geck, A.\,Koitzsch, J.\,Fink, M.\,Knupfer, B.\,Büchner, H.\,Berger, and R.\,Follath, Phys.~Rev.~B \textbf{75}, 172505 (2007);
   \href{http://arxiv.org/pdf/0805.4105}{arXiv:0805.4105} (2008);
   \href{http://arxiv.org/pdf/0807.3929}{arXiv:0807.3929} (2008).
\bibitem{NoteGd} Neutron diffraction measurements on Gd$_2$PdSi$_3$ samples were not possible because of large
absorption. On the other hand, measurements by X-ray resonant magnetic scattering proved not sensitive enough to detect
a possible weak signal from the low-temperature short-range magnetic order in the (110) direction. Preliminary
information about magnetic propagation vector in this material can be found in A.\,Kreyssig, J.-W.\,Kim, L.\,Tan,
D.\,Wermeille, A.\,I.\,Goldman, M.\,Frontzek, and M.\,Löwenhaupt, \textit{Magnetic and chemical superstructures in
Gd$_2$PdSi$_3$ studied using synchrotron radiation}, abstract submitted for the MAR05 meeting of the Amer. Phys. Soc.
(2004).
\bibitem{GondekSzytula07} {\L}.\,Gondek and A.\,Szytu{\l}a, J. Alloys Compd. \textbf{442}, 111 (2007).

\end{thebibliography}
\end{document}